# Equilibrium Shape of Droplets at the Wetted Surface in Strong Gravitational Fields

Leonid Pekker, FujiFilm Dimatix, Inc. Lebanon, New Hampshire 03766, USA
leonid.pekker@fujifilm.com

**Abstract**

In this work, we develop a novel model describing the equilibrium shape of sessile droplets on a wetted horizontal surface in a gravitational field. The model takes into account the intermolecular Lenard-Jones forces between solid and the liquid molecules using the standard disjoining pressure approximation. These forces lead to the formation of a thin, non-removable fluid layer covering the solid substrate. Balancing the disjoining pressure against the surface tension and the gravitational force we calculate the smooth shape of the surface of the liquid. We obtain a criterion when the gravitational forces are so large that they level the droplets completely. We show that, in the case of weak gravitational forces where $\rho\, g\, h^*/\chi \ll 1$, the maximum height of the droplets is described by the classical Quincke's formula $\sqrt{2\,\gamma\,(1-\cos(\theta_e))/\rho\, g}$, where $\gamma$ is the surface tension, $\rho$ is the mass density of the liquid, $g$ is the gravitational constant, $h^*$ is the equilibrium thickness of the non-removable thin liquid film, $\chi$ is the pressure coefficient in the disjoining pressure approximation, and $\theta_e$ is the equilibrium (steady state) contact angle determined by the parameters of the disjoining pressure model and the surface tension; the formula for $\theta_e$ was obtained in the work of L. Pekker, D. Pekker, and N. Petviashvili, "Equilibrium contact angle at the wetted substrate," Phys Fluids 34, 107107 (2022). We also investigate the stability of large droplets when their heights are close to the maximum droplet height.

## I. Introduction

The wetting properties between the liquid and a solid substrate are determined by the cohesive interaction between liquid molecules holding the liquid molecules together and the adhesive interactions between the liquid and the solid molecules [1, 2]. The intermolecular forces can be described by the Lenard-Jones type potentials with a short-range repulsion term and a long-range decaying attraction term. These forces between the liquid and the solid molecules lead to formation of a thin, non-removable fluid film covering the solid substrate. The net effect of these intermolecular potentials on the wetting properties of a liquid film of thickness $h$ can be described by $\gamma$, the surface tension coefficient, and the disjoining pressure $\Pi(h)$, i.e., the net force per unit area of the liquid-solid interface [3-5]. In recent work [6], using the standard disjoining pressure approximation, the authors construct a model describing the shape of sessile droplets on a wetted horizontal substrate in the case of no gravitational field. They also present a formula for the equilibrium (steady state) contact angles for large droplets when the height of the droplet is much larger than the thickness of the non-removable thin fluid film. In [7], the authors show that the formula for the contact angle [6] is applicable for wetted capillaries, slab and cylindrical, and further suggest that this formula is universal regardless of substrate shape.

In this paper, in Section 2, we construct a model for calculating the shape of sessile droplets on a wetted horizontal substrate in a gravitational field. In the model, we assume that the mass density of the fluid above the liquid drop is much smaller than the mass density of the droplet liquid and, therefore, this fluid (such as air) has no effect on the shape of the droplet. As in [6, 7], we use the standard disjoining pressure approximation [3-5]. We obtain a criterion when the gravitational forces are so large that they level the droplets completely. We show that, in the case of weak gravitational forces where $\rho\, g\, h^*/\chi \ll 1$, the maximum height of the droplets is described by the classical Quincke's formula $\sqrt{2\,\gamma\,(1 - \cos(\theta_e))/\rho\, g}$ [8, 9] where $\theta_e$ is the equilibrium (steady state) contact angle determined by the parameters of the disjoining pressure model and the surface tension [6]. In this section, we also investigate the stability of large droplets when their heights approach the maximum droplet height. In Section 3, we compare the droplet shapes calculated by the full model derived in Section 2 and the reduced (classical) model in which the disjoining pressure is dropped. We show that when $\rho g h^*/\chi \ll 1$, the droplet shapes calculated by both models are identical not only for extremely large droplets, but also for droplets as small as $h_{max} > 20h^*$ for both small and large contact angles. Concluding remarks are given in Section 4.

## II. Model of steady-state droplet placed on solid horizontal substrate in a gravitational field

Let us consider the steady-state shape of a droplet in a gravitational field placed on a horizontal substrate that supports a non-removable thin liquid film. In the model, for the sake of simplicity, we assume that the droplet is invariant with respect to translation along the y-axis, Fig. 1. Then, the equation describing the shape of $x = h(z)$ of the droplet, Fig. 1, can be written as

$$-\frac{\gamma \frac{d^2 h}{dz^2}}{\left(1+\left(\frac{dh}{dz}\right)^2\right)^{1.5}} - \chi\left\{\left(\frac{h^*}{h}\right)^m - \left(\frac{h^*}{h}\right)^n\right\} + \rho g h = p. \tag{1}$$

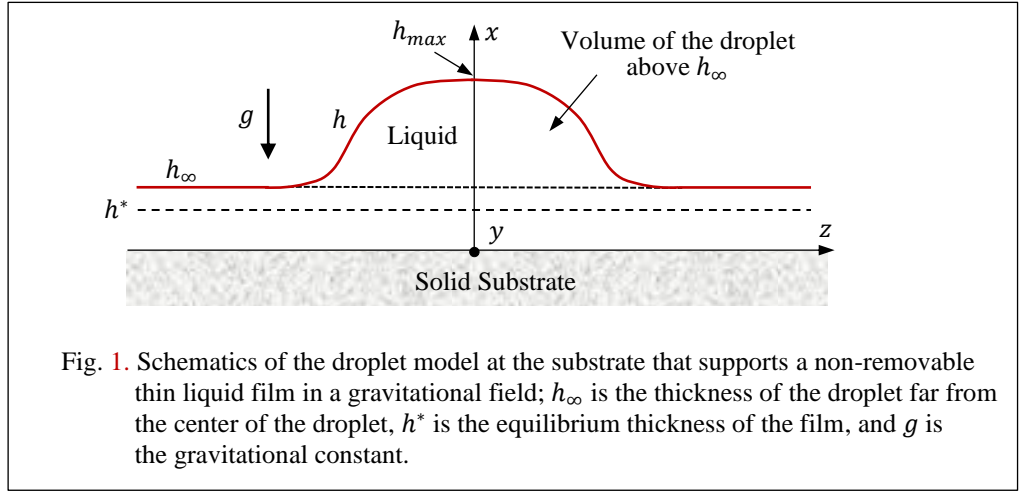

Fig. 1. Schematics of the droplet model at the substrate that supports a non-removable thin liquid film in a gravitational field; $h_\infty$ is the thickness of the droplet far from the center of the droplet, $h^*$ is the equilibrium thickness of the film, and $g$ is the gravitational constant.

Eq. (1) states that the pressure at the free surface $p$ is determined by the balance of the surface tension pressure, the disjoining pressure associated with the intermolecular force model parameters $\chi = A/(h^*)^3$, $h^*$, $m$, and $n$ of Ref. [10], and the gravitational pressure, correspondingly the first, second, and third terms in the LHS of Eq. (1). Here, A is the Hamaker constant, and $m$ and $n$ parametrize the dependence of the disjoining pressure on film thickness. The model assumes that the mass density of a fluid (such as air) above the liquid drop is much smaller than the mass density of the liquid and, therefore, has no effect on the shape of the droplet. To ensure the stability of the non-removable film, we assume that $m > n$. This assumption corresponds to the Lenard-Jones intermolecular type potential, where the molecules repel each other when the distance between them is small and attract each other when the distance between them is large. As shown in [11], integrating the Lenard-Jones type potential, $V(r) = \epsilon\left((\sigma/r)^k - (\sigma/r)^l\right)$, over the volume of the solid gives the disjoining pressure with power constants $m = k - 3$ and $n = l - 3$.

Introducing the dimensionless variables $Z$, $H$, $P$ and the dimensionless lubrication and gravitational parameters $\alpha_l$ and $\alpha_g$, as

$$z = h^*Z, \qquad h = h^*H, \qquad p = \frac{h^*}{\gamma}P, \qquad \alpha_l = \frac{\chi h^*}{\gamma}, \qquad \alpha_g = \frac{\rho g (h^*)^2}{\gamma}, \qquad (2)$$

Eq. (1) becomes:

$$-\frac{\frac{d^2 H}{dZ^2}}{\left(1+\left(\frac{dH}{dZ}\right)^2\right)^{1.5}} - \alpha_l\left(\frac{1}{H^m} - \frac{1}{H^n}\right) + \alpha_g H = P. \qquad (3)$$

Fig. 2 shows the dimensionless shape of the droplet shape $X = H(Z)$, which is illustrated in dimension variables in Fig. 1.

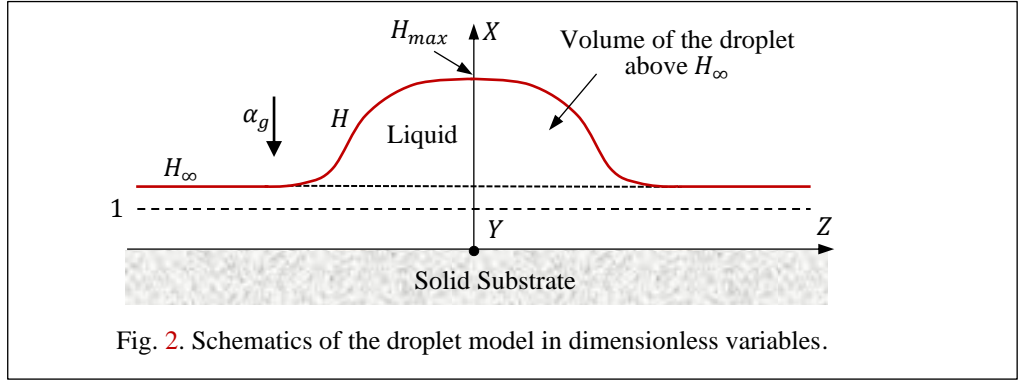

Fig. 2. Schematics of the droplet model in dimensionless variables.

Taking into account that far from the droplet $H(Z) \underset{Z \to \pm\infty}{\to} H_\infty$ and $\frac{dH}{dZ}, \frac{d^2 H}{dZ^2} \underset{Z \to \pm\infty}{\to} 0$ (see Fig. 2), we obtain,

$$P = -\alpha_l\left(\frac{1}{H_\infty{}^m} - \frac{1}{H_\infty{}^n}\right) + \alpha_g H_\infty \qquad (4)$$

and

$$-\frac{\frac{d^2H}{dZ^2}}{\alpha_l\left(1+\left(\frac{dH}{dZ}\right)^2\right)^{1.5}} - \left(\frac{1}{H^m} - \frac{1}{H^n}\right) + \frac{\alpha_g}{\alpha_l}H = -\left(\frac{1}{H_\infty{}^m} - \frac{1}{H_\infty{}^n}\right) + \frac{\alpha_g}{\alpha_l}H_\infty, \tag{5}$$

where $H_\infty$ is the thickness of the film far from the droplet (see Fig. 2).

Let us analyze Eqs. (4) and (5) to determine the maximum values of $H_\infty$ due to the gravitational force. Taking the derivative with respect to $H_\infty$ from Eq. (4),

$$\frac{1}{\alpha_l}\frac{dP}{dH_\infty} = \frac{m}{H_\infty{}^{m+1}} - \frac{n}{H_\infty{}^{n+1}} + \frac{\alpha_g}{\alpha_l}, \tag{6}$$

and introducing the function

$$y = \frac{m}{x^{m+1}} - \frac{n}{x^{n+1}} \tag{7}$$

and parameter

$$\beta = \frac{\alpha_g}{\alpha_l} = \frac{\rho\, g\, h^*}{\chi}, \tag{8}$$

we obtain that:

(a) $y(x)$ crosses the $x$-axis at $x_1 = \sqrt[m-n]{m/n}$, Fig. 3;

(b) $y(x)$ reaches its minimum value $y_2$ at $x_2 = \sqrt[m-n]{m(m+1)/n(n+1)}$, Fig. 3;

(c) for $\beta > \beta_{cr}$,

$$\beta_{cr} = -y_2 = n\left(\frac{n(n+1)}{m(m+1)}\right)^{\frac{n+1}{m-n}} - m\left(\frac{n(n+1)}{m(m+1)}\right)^{\frac{m+1}{m-n}} \tag{9}$$

$P$ is monotonically increases with an increase in $H_\infty$, Eq. (6).

This means that first, in the case of $\beta = 0$, i.e., no gravitational field, $H_{\infty,max}=\sqrt[m-n]{m/n}$ [6]; and second, when $\beta > \beta_{cr}$, i.e., the case of a strong gravitational field, Eq. (5) does not have a "droplet" form solution as might be expected from Fig. 2. Indeed, as one can see from Eq. (5), for $\beta > \beta_{cr}$, when $H > H_\infty$, $d^2H/dZ^2$ in the left-hand side of Eq. (5) is always positive – this is nonsensical because at the top of the droplet, at $H = H_{max}$, $d^2H/dZ^2$ should be negative, see Fig. 2. Thus, we have demonstrated that when $\beta \geq \beta_{cr}$, the gravitational field is so large that it flattens the droplets completely. In Fig. 4, we show $H_{\infty,max}$ vs. $\beta$.

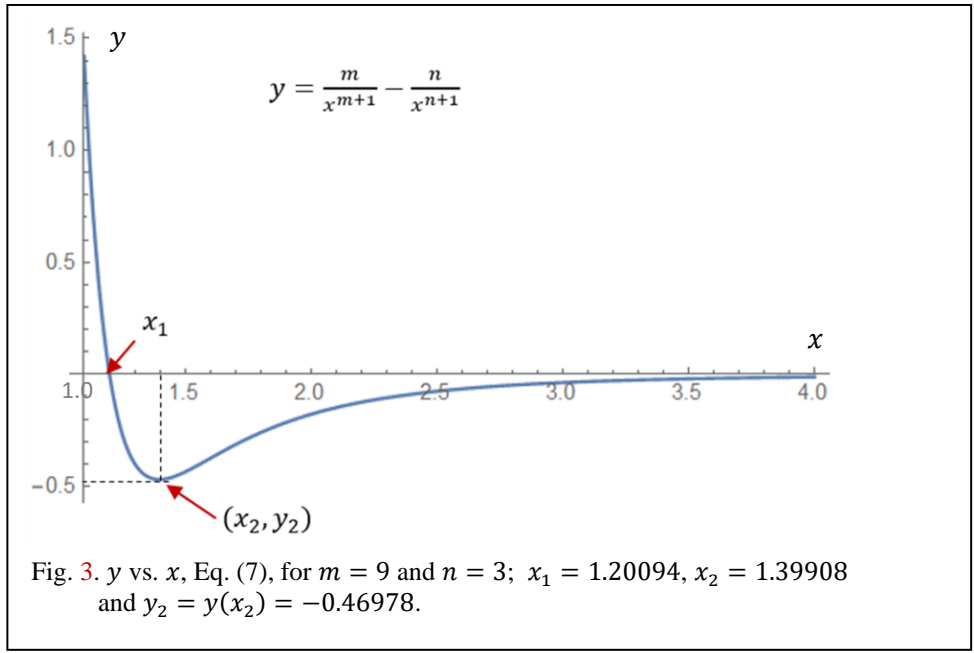

Fig. 3. $y$ vs. $x$, Eq. (7), for $m = 9$ and $n = 3$; $x_1 = 1.20094$, $x_2 = 1.39908$ and $y_2 = y(x_2) = -0.46978$.

It is worth noting that according to Eqs. (2) and (8), $\beta$ is the ratio of the gravitational pressure at the equilibrium thickness of the film, $\rho g h^*$, to the characteristic value of disjoining pressure, $\chi$. That is why, when $\beta > \beta_{cr}$, the gravitational forces are so strong that they level the droplets completely to a flat-thin-liquid-film with the thickness slightly larger than $h^*$; as we will see later (Fig 5), at $\beta = \beta_{cr}$, the thickness of this film is 1.39909 which corresponds to $H_{\infty,max}(\beta_{cr})$ in Fig. 4.

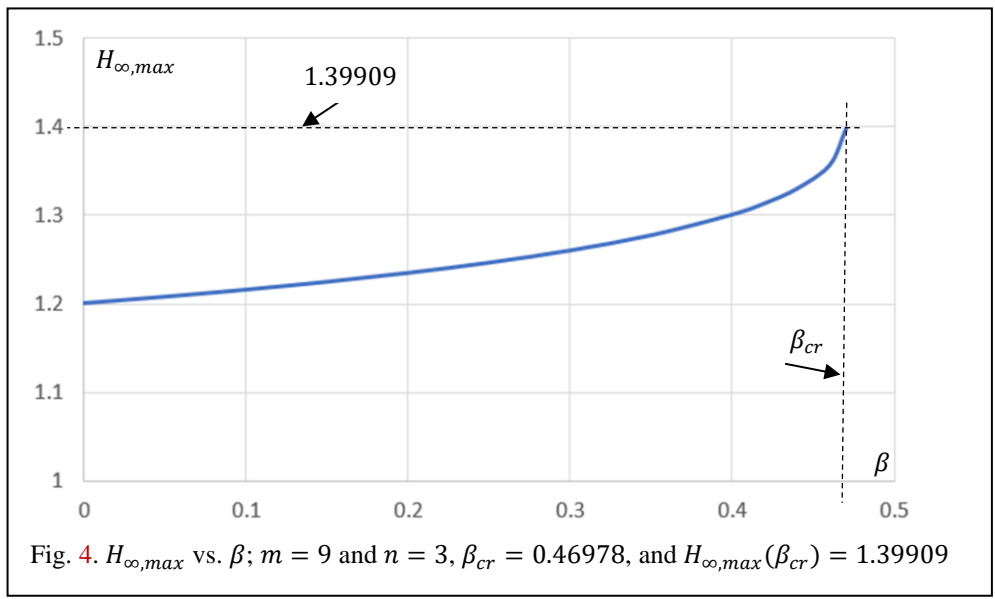

Fig. 4. $H_{\infty,max}$ vs. $\beta$; $m = 9$ and $n = 3$, $\beta_{cr} = 0.46978$, and $H_{\infty,max}(\beta_{cr}) = 1.39909$

Now let us integrate Eq. (5). Multiplying Eq. (5) by $\frac{dH}{dZ}$ and integrating the resulting equation, we obtain:

$$-\frac{\frac{dH}{dZ}\frac{d^2H}{dZ^2}}{\alpha_l\left(1+\left(\frac{dH}{dZ}\right)^2\right)^{1.5}} - \left(\frac{1}{H^m} - \frac{1}{H^n}\right)\frac{dH}{dZ} + \beta H \frac{dH}{dZ} = \left(-\left(\frac{1}{H_\infty{}^m} - \frac{1}{H_\infty{}^n}\right) + \beta H_\infty\right)\frac{dH}{dZ} \rightarrow$$

$$\rightarrow \frac{d}{dZ}\left(\frac{1}{\alpha_l\left(1+\left(\frac{dH}{dZ}\right)^2\right)^{0.5}} - \left(-\frac{1}{(m-1)H^{m-1}} + \frac{1}{(n-1)H^{n-1}}\right) + \frac{\beta}{2}H^2\right) =$$

$$= \frac{d}{dZ}\left(H\left(-\left(\frac{1}{H_\infty{}^m} - \frac{1}{H_\infty{}^n}\right) + \beta H_\infty\right)\right) \rightarrow$$

$$\rightarrow \frac{1}{\alpha_l\left(1+\left(\frac{dH}{dZ}\right)^2\right)^{0.5}} = \left(-\frac{1}{(m-1)H^{m-1}} + \frac{1}{(n-1)H^{n-1}}\right) - \frac{\beta}{2}H^2 + H\left(-\left(\frac{1}{H_\infty{}^m} - \frac{1}{H_\infty{}^n}\right) + \beta H_\infty\right) + C, \quad (10)$$

where $C$ is a constant of integration. Taking into account that $\left.\frac{dH}{dZ}\right|_{Z=\pm\infty} = 0$, Fig. 2, we obtain from Eq. (10) that

$$C = \frac{1}{\alpha_l} + \left(\frac{m}{(m-1)H_\infty{}^{m-1}} - \frac{n}{(n-1)H_\infty{}^{n-1}}\right) - \frac{\beta}{2}H_\infty{}^2. \quad (11)$$

Substituting Eq. (11) into Eq. (10) and squaring the resulting equation, we can present Eq. (10) in the following form:

$$\left(\frac{dH}{dZ}\right)^2 = \frac{B-(0.5B)^2}{(1-0.5B)^2}, \quad (12)$$

$$B = 2\alpha_l \begin{pmatrix} H\left(\left(\frac{1}{H_\infty{}^m} - \frac{1}{H_\infty{}^n}\right) - \beta H_\infty\right) + \left(\frac{1}{(m-1)H^{m-1}} - \frac{1}{(n-1)H^{n-1}}\right) + \\ +\frac{\beta}{2}H^2 - \left(\frac{m}{(m-1)H_\infty{}^{m-1}} - \frac{n}{(n-1)H_\infty{}^{n-1}}\right) + \frac{\beta}{2}H_\infty{}^2 \end{pmatrix}. \quad (13)$$

In our simulation, we use the boundary condition $H(0) = H_{max}$.

Next, let us determine $H_{MAX}$, the maximum value of $H_{max}$ at a given $\beta$, vs. $\beta$. In the case of no gravitational field, $\beta = 0$, the height of the droplet is not limited [6]. However, in a gravitational field, $H_{max}$ is limited. Indeed, since the pressure in the droplet is positive, the right-hand side of Eq. (5) is positive; and, since at $H_{max} \rightarrow \infty$, we can drop the small disjoining pressure term in the left-hand side of Eq. (5), we obtain from Eq. (5) that $(d^2H/dZ^2)_{H=H_{max}\rightarrow\infty} > 0$; this makes no physical sense. Thus, we have shown that $H_{MAX}$ corresponds to $H_{max}$ at which $(d^2H/dZ^2)_{H=H_{max}} = 0$. Taking this into account, we obtain the following set of equations for $H_{MAX}$ and corresponding $H_\infty$:

$$H_{MAX}\left(\left(\frac{1}{H_\infty{}^m} - \frac{1}{H_\infty{}^n}\right) - \beta H_\infty\right) + \left(\frac{1}{(m-1)H_{MAX}{}^{m-1}} - \frac{1}{(n-1)H_{MAX}{}^{n-1}}\right) +$$

$$+ \frac{\beta}{2}H_{MAX}{}^2 - \left(\frac{m}{(m-1)H_\infty{}^{m-1}} - \frac{n}{(n-1)H_\infty{}^{n-1}}\right) + \frac{\beta}{2}H_\infty{}^2 = 0 \quad (14)$$

$$-\left(\frac{1}{H_{MAX}{}^m} - \frac{1}{H_{MAX}{}^n}\right) + \beta\, H_{MAX} = -\left(\frac{1}{H_\infty{}^m} - \frac{1}{H_\infty{}^n}\right) + \beta H_\infty. \quad (15)$$

Eq. (14) follows from Eqs. (12) and (13) and the assumption that $(dH/dZ)_{H=H_{max}} = 0$. Similarly, Eq. (15) follows from Eq. (5) and the assumption that $(d^2H/dZ^2)_{H=H_{max}} = 0$. The solution of this set of equations for the case of $m = 9$ and $n = 3$ is presented in Fig. 5. As expected, $(H_{MAX})_{\beta \to \beta_{cr}} = (H_\infty)_{\beta \to \beta_{cr}} = x_2$. For $\beta > \beta_{cr}$, this system of equations has a trivial solution: $H_{MAX} = H_\infty$. The case of $(H_{MAX})_{\beta \to 0} \to \infty$ and $(H_\infty)_{\beta \to 0} \to 1$ corresponds to the case of no gravity [6].

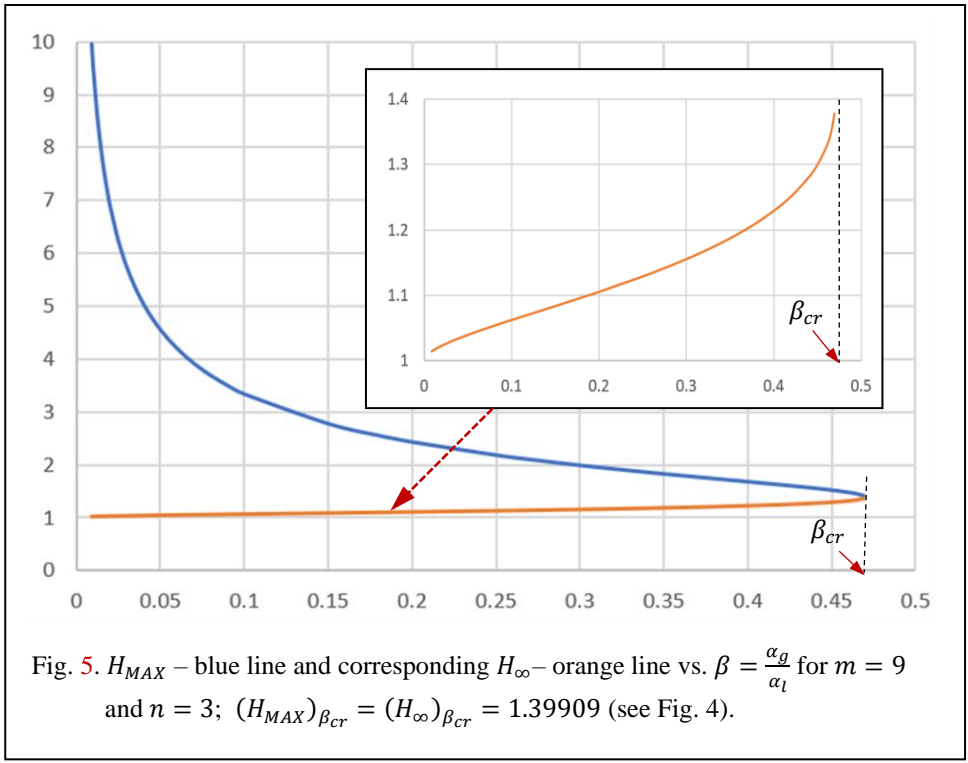

Fig. 5. $H_{MAX}$ – blue line and corresponding $H_\infty$ – orange line vs. $\beta = \frac{a_g}{a_l}$ for $m = 9$ and $n = 3$; $(H_{MAX})_{\beta_{cr}} = (H_\infty)_{\beta_{cr}} = 1.39909$ (see Fig. 4).

Now let us obtain a formula for $H_{MAX}$ in the case of a weak gravitational field when $\beta \ll 1$. In weak gravitational fields, $H_{MAX} \gg 1$ and $H_\infty - 1 \ll 1$. Writing $H_\infty = 1 + \varepsilon$, where $\varepsilon \ll 1$, the set of Eqs. (14) and (15) reduces to the following form:

$$-H_{MAX}(m-n)\varepsilon + \frac{\beta}{2} H_{MAX}^2 + \frac{m-n}{(m-1)(n-1)} = 0, \tag{16}$$

$$\beta H_{MAX} = (m-n)\varepsilon. \tag{17}$$

Here, in Eq. (14), in the first parentheses, we have approximated $(1/(1+\varepsilon)^m - 1/(1+\varepsilon)^n)$ as $-(m-n)\varepsilon$ and dropped $\beta H_\infty$; we dropped both terms in the second parentheses; in the third parentheses, in both terms, we substituted unity for $H_\infty$; and we dropped the last term in the LHS of this equation. In Eq. (15), in the LHS, we have dropped both terms in the parentheses; and, in the RHS, in the parentheses we approximated the terms as $(m-n)\varepsilon$ and dropped the last term. Solving this system of equations for $H_{MAX}$, we obtain

$$H_{MAX} = \sqrt{\frac{2(m-n)}{\beta(m-1)(n-1)}}. \tag{18}$$

As expected, $(H_{MAX})_{\beta \to 0} \to \infty$. Substituting $\beta$ from Eq. (8) into Eq. (18) and then using Eqs. (2), we obtain the maximum height of the droplet in dimensional form:

$$h_{MAX} = \sqrt{\frac{2(m-n)\chi h^*}{(m-1)(n-1)\rho g}}. \tag{19}$$

Now let us show that the Eq. (19) agrees with the classical formula of G. H. Quincke [8, 9]. Since, in weak gravitational fields, the equilibrium contact angle for large droplets is independent of the gravitational constant, we can use the formula contact angle [6],

$$\tan(\theta_e) = \frac{\sqrt{\frac{2(m-n)\chi h^*}{(m-1)(n-1)\gamma} - \left(\frac{(m-n)\chi h^*}{(m-1)(n-1)\gamma}\right)^2}}{1 - \frac{(m-n)\chi h^*}{(m-1)(n-1)\gamma}}. \tag{20}$$

Substituting $(m-n)\chi h^*/(m-1)(n-1)$ from Eq. (20) into Eq. (19), we obtain the G. H. Quincke's formula [8, 9], see Appendix A,

$$h_{MAX} = \sqrt{\frac{2\gamma(1-\cos(\theta_e))}{\rho g}}. \tag{21}$$

Fig. 6. shows $H_{MAX}$ and $H_{MAX} - H_\infty$ vs. $\beta$. $H_{MAX}$ is calculated using the assumption of a weak gravitational field, i.e., the reduced model as described in Eq. (18); $H_{MAX} - H_\infty$ is calculated using the set of Eqs. (15) and (16): the full model. In this plot, to make the two approaches directly comparable, we have used $H_{MAX} - H_\infty$ to eliminate the shift $H_\infty$ in the full model, Fig. 2. As one can see, $(H_{MAX} - H_\infty)_{\beta_{cr}} = 0$, which corresponds to Fig. 6. As expected, with a decrease in $\beta$, the differences between the models decrease.

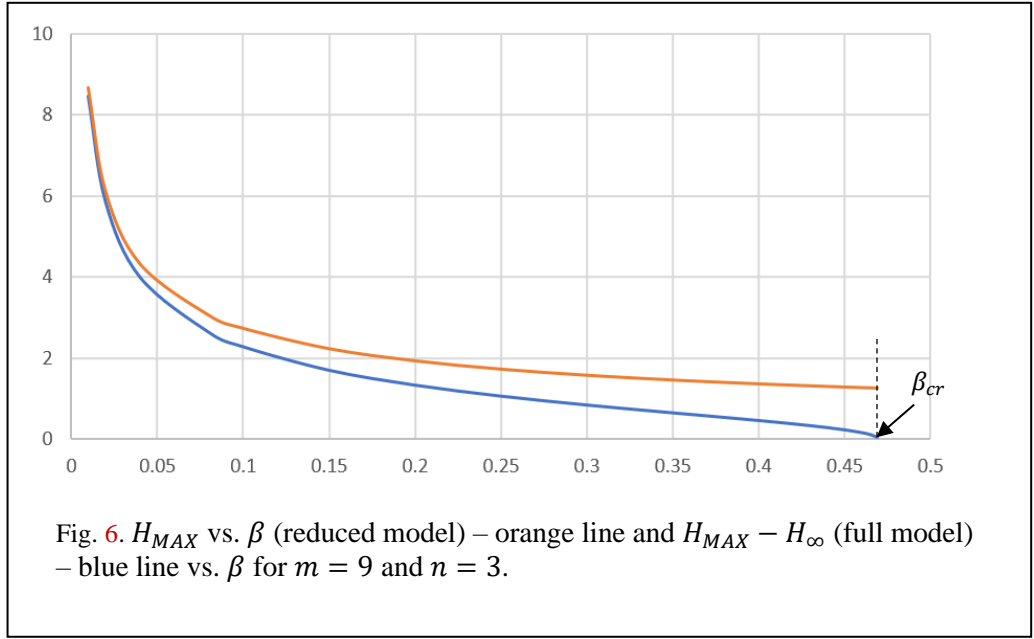

Fig. 6. $H_{MAX}$ vs. $\beta$ (reduced model) – orange line and $H_{MAX} - H_\infty$ (full model) – blue line vs. $\beta$ for $m = 9$ and $n = 3$.

Now, let us consider the stability of large droplets when their heights approach $H_{MAX}$. Since all odd derivatives of $H$ at $Z = 0$ are zero (because of the symmetry of the droplet with respect to the $X$-axis, Fig. 2), all even derivatives of $H$ at $H = H_{max}$ can be written as

$$\left(\frac{d^{2k}H}{dZ^{2k}}\right)_{H=H_{max}} = F_k\left(f_{k,k-1}(H)\left(\frac{d^{2(k-1)}H}{dZ^{2(k-1)}}\right), f_{k,k-2}(H)\left(\frac{d^{2(k-2)}H}{dZ^{2(k-2)}}\right), \ldots, f_{k,k-1}(H)\left(\frac{d^{2}H}{dZ^{2}}\right)\right)_{H=H_{max}} \quad (22)$$

where $k = 2, 3, 4 \ldots$ The explicit expressions for functions $F_k$ and $f_{k,k-1}$ can be obtained from Eq. (5). Thus, since $(d^2H/dZ^2)_{H=H_{MAX}} = 0$, we obtain from Eqs. (22) that when $H_{max} \to H_{MAX}$, all even derivatives of $H$ at $Z = 0$ tend to zero as well, $(d^{2k}H/dZ^{2k})_{H=H_{max}} \to 0$. This means that with an increase in the volume of the droplet, the droplet extends in the $Z$-direction, becoming a flat uniform layer with its thickness equal to $H_{MAX}$ everywhere except for the droplet "sides", Fig. 7. Therefore, we can determine the stability of large droplets with $H_{max} \to H_{MAX}$ by approximating them as a uniform liquid layer with thickness $H_{MAX}$.

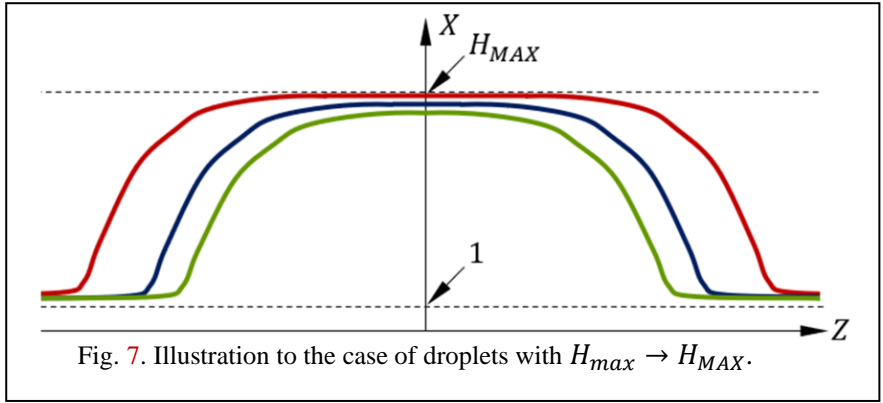

Fig. 7. Illustration to the case of droplets with $H_{max} \to H_{MAX}$.

We consider the linear stability of a flat liquid layer using the standard lubrication approximation [10]. The governing equation describing the dynamic of the liquid layer [10],

$$\frac{\partial h}{\partial t} = -\frac{\partial}{\partial z}\left(\frac{h^3}{3\mu}\frac{\partial}{\partial h}\left(\gamma\frac{\partial^2 h}{\partial z^2} + \chi\left\{\left(\frac{h^*}{h}\right)^m - \left(\frac{h^*}{h}\right)^n\right\} - \rho g h\right)\right), \qquad (23)$$

can be presented in dimensionless form as

$$\frac{\partial H}{\partial T} = \frac{\partial}{\partial Z}\left(H^3 \frac{\partial}{\partial Z}\left(-\frac{1}{\alpha_l}\frac{\partial^2 H}{\partial Z^2} - \left(\frac{1}{H^m} - \frac{1}{H^n}\right) + \beta H\right)\right), \qquad (24)$$

$$t = \frac{3\mu}{\chi}T. \qquad (25)$$

Here, $\mu$ is the viscosity of the liquid, $t$ is time, and $T$ is dimensionless time. Linearizing Eq. (24) with respect to the not perturbed liquid layer with uniform thickness $H_{MAX}$, $H = H_{MAX} + \delta H e^{\Gamma t + iKZ}$, we obtain the following dispersion equation describing the stability of the layer:

$$\Gamma e^{iKZ} = (H_{MAX})^3 \frac{\partial}{\partial Z}\left(\frac{\partial}{\partial Z}\left(-\frac{1}{\alpha_l}\frac{\partial^2}{\partial Z^2}(e^{iKZ}) + \left(\frac{m}{(H_{MAX})^{m+1}}e^{iKZ} - \frac{n}{(H_{MAX})^{n+1}}e^{iKZ}\right) + \beta e^{iKZ}\right)\right) \to$$

$$\to \Gamma = (H_{MAX})^3 K^2 \left(-\frac{1}{\alpha_l}K^2 - \frac{m}{(H_{MAX})^{m+1}} + \frac{n}{(H_{MAX})^{n+1}} - \beta\right). \qquad (26)$$

Let us introduce $\phi(\beta)$ as

$$\phi(\beta) = -\frac{1}{\beta}\frac{m}{(H_{MAX})^{m+1}} + \frac{1}{\beta}\frac{n}{(H_{MAX})^{n+1}} - 1. \qquad (27)$$

Then, as follows from Eq. (26), if $\phi(\beta) < 0$, the "flat" droplets are stable, and if $\phi(\beta) > 0$, the "flat" droplets are unstable. In weak gravitational fields, $\beta \ll 1$, substituting $H_{MAX}$ from Eq. (18) into Eq. (27) we obtain that for $m = 9$ and $n = 3$, $\phi(\beta \to 0) \to -1$, i.e., flat droplets in weak gravitational fields are stable. At $\beta = \beta_{cr}$, $H_{MAX} = H_\infty = x_2 = \sqrt[m-n]{m(m+1)/n(n+1)}$, Fig. 3, and as follows from Eq. (9), $\phi(\beta_{cr}) = 0$. As one can see from Fig. 8, $\phi(\beta) < 0$ whenever $\beta < \beta_{cr}$. Thus, we have shown that in gravitational fields, large volume droplets with $H_{max} \to H_{MAX}$ are always stable.

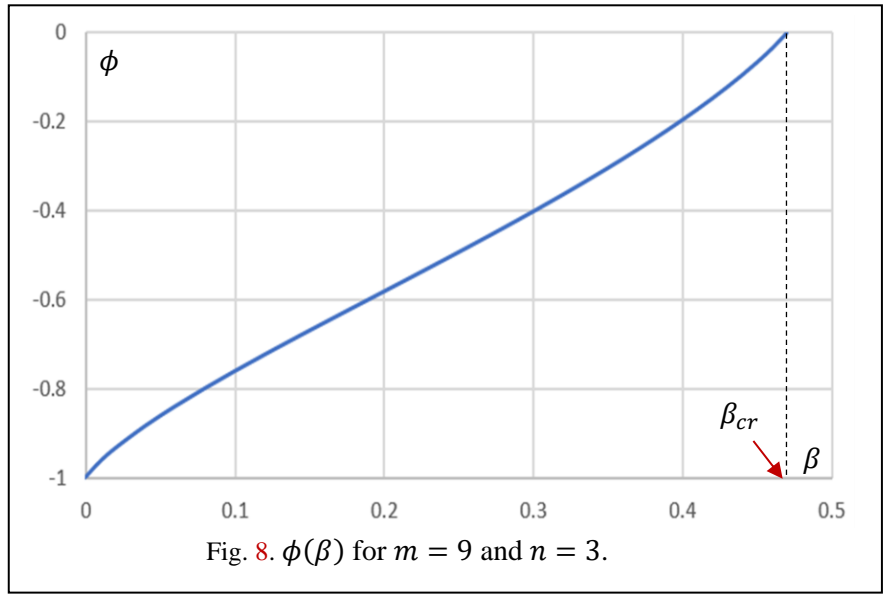

Fig. 8. $\phi(\beta)$ for $m = 9$ and $n = 3$.

### III. Comparison of the shapes of the droplets calculated via the full and reduced (classical) models

In this section, we consider a reduced dimensionless droplet model, in which we drop the disjoining pressure terms, Eq. (28), and assume that the contact angle is specified by Eq. 20 [6]; this is the classical Quincke model. The full model is described in Section II, Eqs. (12) and (13).

$$-\frac{\frac{d^2H}{dZ^2}}{\left(1+\left(\frac{dH}{dZ}\right)^2\right)^{1.5}} + \alpha_g H = P. \tag{28}$$

The aim of this section is to compare the reduced model against the full model.

Let us obtain an equation describing the shape of the droplet calculated by the reduced model with a given droplet height. Multiplying Eq. (28) by $\frac{dH}{dZ}$ and integrating the resulting equation, we obtain:

$$\frac{1}{\sqrt{1+\left(\frac{dH}{dZ}\right)^2}} + \alpha_g \frac{H^2}{2} - PH = C, \tag{29}$$

where $C$ is an integration constant. Since $(dH/dZ)_{H=H_{max}} = 0$ (see Fig. 2), we obtain from Eq. (29) that

$$C = 1 + \alpha_g \frac{(H_{max})^2}{2} - PH_{max}. \tag{30}$$

Substituting Eq. (30) into Eq. (29), we obtain:

$$\frac{1}{\sqrt{1+\left(\frac{dH}{dZ}\right)^2}} + \alpha_g \frac{H^2}{2} - PH = 1 + \alpha_g \frac{(H_{max})^2}{2} - PH_{max}. \tag{31}$$

At $H_{max} = H_{MAX}$, as follows from Eq. (28), $(P)_{H_{MAX}} = \alpha_g H_{MAX}$ because of $(d^2H/dZ^2)_{H_{MAX}} = 0$. Further, as follows from Eq. (31), $(1 + ((dH/dZ)_{H=0})^2)^{-1/2} = 1 + \alpha_g \frac{(H_{MAX})^2}{2} - H_{MAX}(P)_{H_{MAX}}$. Therefore, at $H_{max} = H_{MAX}$ obtain that

$$\left(1 + \left(\left(\frac{dH}{dZ}\right)_{H=0}\right)^2\right)^{-1/2} = 1 - \alpha_g \frac{(H_{MAX})^2}{2}. \tag{32}$$

Substituting Eq. (32) into Eq. (31) at $H = 0$, we obtain an equation for the pressure

$$P = \alpha_g \frac{(H_{MAX})^2 + (H_{max})^2}{2 H_{max}}. \tag{33}$$

Substituting Eq. (33) into Eq. (31), we obtain the final equation describing the shape of the droplets in the framework of the reduced model,

$$\frac{1}{\sqrt{1+\left(\frac{dH}{dZ}\right)^2}} + \alpha_g \frac{H^2}{2} - H\alpha_g \left(\frac{(H_{MAX})^2+(H_{max})^2}{2 H_{max}}\right) = 1 + \alpha_g \frac{(H_{max})^2}{2} - \alpha_g \frac{(H_{MAX})^2+(H_{max})^2}{2 H_{max}} H_{max} \rightarrow$$

$$\rightarrow \frac{1}{\sqrt{1+\left(\frac{dH}{dZ}\right)^2}} 1 + \alpha_g \frac{(H_{max})^2-H^2}{2} - \alpha_g \frac{(H_{MAX})^2+(H_{max})^2}{2 H_{max}} (H_{max} - H) \rightarrow$$

$$\rightarrow \frac{1}{\sqrt{1+\left(\frac{dH}{dZ}\right)^2}} = 1 + \alpha_g \frac{(H_{max})^2-H^2}{2} - \alpha_g \frac{(H_{MAX})^2+(H_{max})^2}{2 H_{max}} (H_{max} - H). \tag{34}$$

Solving Eq. (34) for $dH/dz$ we obtain,

$$\left(\frac{dH}{dZ}\right)^2 = \frac{b-(0.5b)^2}{(1-0.5b)^2}, \tag{35}$$

$$b = \alpha_g\left(-H^2 - (H_{max})^2 + (H_{max} - H)\frac{(H_{MAX})^2+(H_{max})^2}{H_{max}}\right). \tag{36}$$

As we can see, the form of Eq. (35) is the same as Eq. (12), but with a different function $b$. In our simulation, we use the boundary condition at $Z = 0$, $H(0) = H_{max}$.

Since $H_{MAX}$, the maximum height of the droplets, and corresponding $H_\infty$ are governed by the parameter $\beta$, see Fig. 5, in Figs. 9 and 10, we compare the reduced model against the full model for $\beta = 0.1, 0.01$, and $0.001$. In these simulations, we took $m = 9$ and $n = 3$; $\alpha_l = 0.5$ which corresponds to an effective $\theta_e = 151.0°$ (Fig. 9) and $\alpha_l = 0.3$ which corresponds to an effective $\theta_e = 27.4°$ (Fig. 10); and used different values of $H_{max}$. In these figures, for the full model, to eliminate the shift of the droplets due to the non-removable thin liquid film, Fig. 2, the height of the droplet was taken as $H_{max} - H_\infty$, as we did in Fig. 6. In the case of $\beta = 0.1$, Fig. 9a and 10a, where $H_{MAX} - H_\infty = 2.2813$, we see large differences between the models for both values of $\alpha_l$, i.e., for small and large apparent contact angles. This was expected because the reduced model, Eqs. (20), (35) and (36), work reasonably well only for much larger droplets, i.e., when $H_{max} \gg 1$. With a decrease in $\beta$, the "window" for $H_{max}$ increases. Therefore, the differences between the models in Figs. 9b and 10b where $\beta = 0.01$ and $H_{MAX} - H_\infty = 8.4714$, are smaller than in Figs. 9a and 10a; and, in Figs. 8c and 9c where $\beta = 0.001$ and $H_{MAX} - H_\infty = 27.32$, they are smaller than in Figs. 9b and 10b. Thus, we have demonstrated that except for the transition region of a few $h^*$ above the substrate, where, in our model, the shape of the droplet transfers into the plateau (see Figs. 9 and 10), with an increase in the height of droplet, the differences between the shapes of the droplet calculated by the two models decrease and become negligibly small when $H_{max} \gg 1$.

IV. **Concluding Remarks**

In this paper, we extended the steady-state model describing the equilibrium shape of sessile droplets on a wetted surface [6] to the case of a gravitational field. In the model, we assumed that the substrate is horizontal with respect to the vertical gravitational force and covered by a non-removable thin liquid film.

In the model, we described the intermolecular interaction between the solid and liquid model using the classical disjoining pressure approximation. This allowed us to obtain a smooth fluid surface of the droplet along the substrate; at infinity, the fluid surface is a flat thin film with thickness that is slightly larger than the equilibrium thickness of the film and determined by the pressure in the droplet. We show that in gravitational fields, large volume droplets with $H_{max} \to H_{MAX}$ are always stable.

We have also derived a condition at which the gravitational pressure overpowers the disjoining pressure and flattens the droplets completely. We have shown, that when $\beta > \beta_{cr}$, the set of Eqs. (12) and (13) describing the shape of the droplet does not have a droplet-type solution, only a uniform-thickness-film solution. It should be noted that for typical liquids like water with $\rho = 10^3 \text{kg/m}^3$, $h^* = 10^{-9}$ m, $\chi = 10^6$ Pa [10] in the gravitational field of the Earth, $g = 10 m/sec^2$, then $\beta = 10^{-11}$ which is many orders smaller than $\beta_{cr}$. Thus, the obtained criterion likely has only theoretical value.

We also compared our model against Quincke's classical model [8,9] in which the disjoining pressure is ignored, and the contact angle is assumed to be known and independent of gravitation; in our simulation we have used the Pekker-Pekker-Petviashvilli formula for the contact angle [6]. We have demonstrated that for both small and large contact angles, the differences in the shapes of the droplet calculated by our and Quincke's models becomes negligibly small when the height of the droplets is larger than $20h^*$, which corresponds to $\beta$ smaller than 0.001. However, the differences between the models are large in the transition region of a few $h^*$ above the substrate, where the shape of the droplet, in our model, transfers into the plateau, and, in Quincke's model, it crosses the substrate at the apparent equilibrium contact angle $\theta_e$.

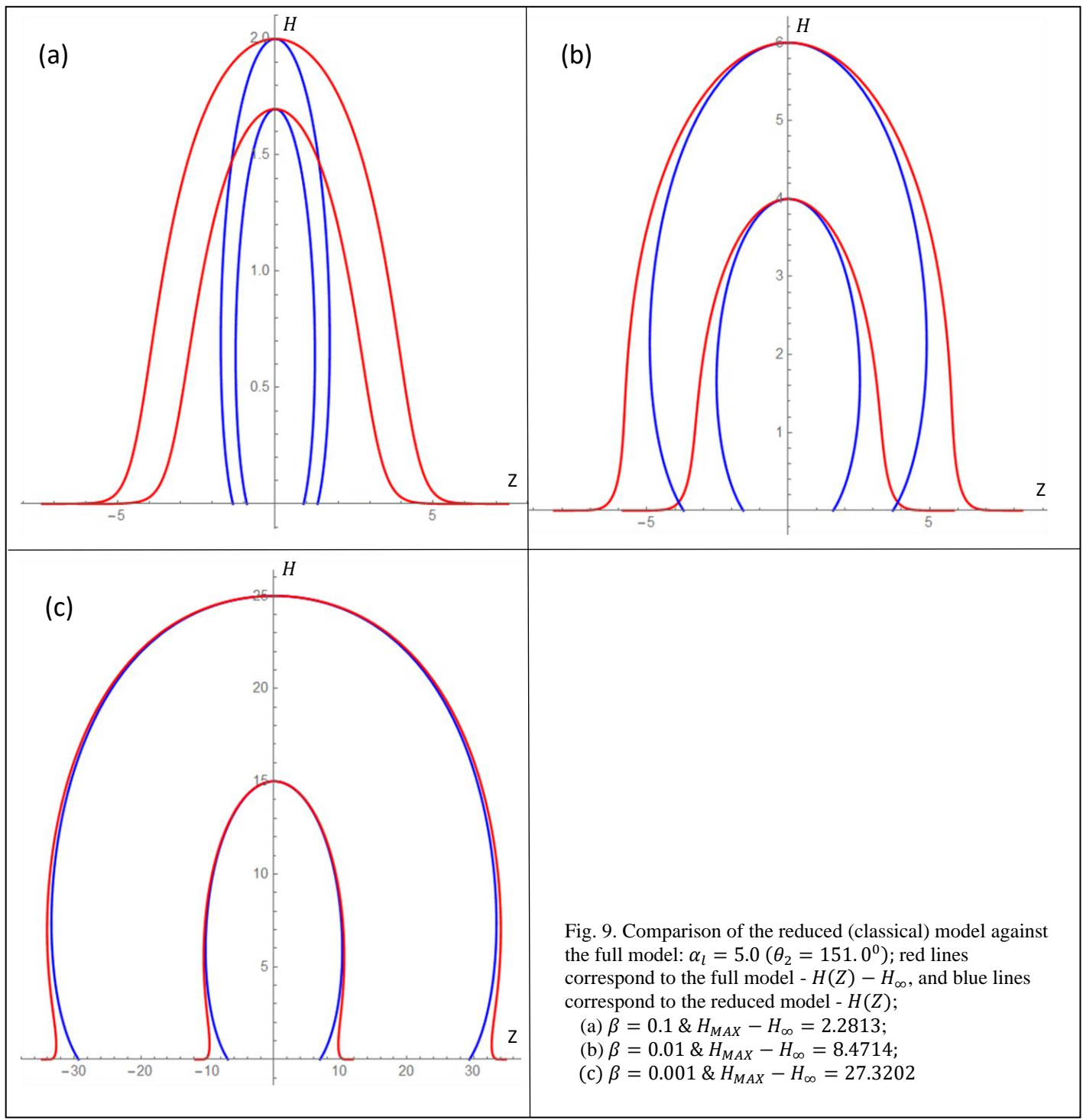

Fig. 9. Comparison of the reduced (classical) model against the full model: $\alpha_l = 5.0$ ($\theta_2 = 151.0^0$); red lines correspond to the full model - $H(Z) - H_\infty$, and blue lines correspond to the reduced model - $H(Z)$;
(a) $\beta = 0.1$ & $H_{MAX} - H_\infty = 2.2813$;
(b) $\beta = 0.01$ & $H_{MAX} - H_\infty = 8.4714$;
(c) $\beta = 0.001$ & $H_{MAX} - H_\infty = 27.3202$

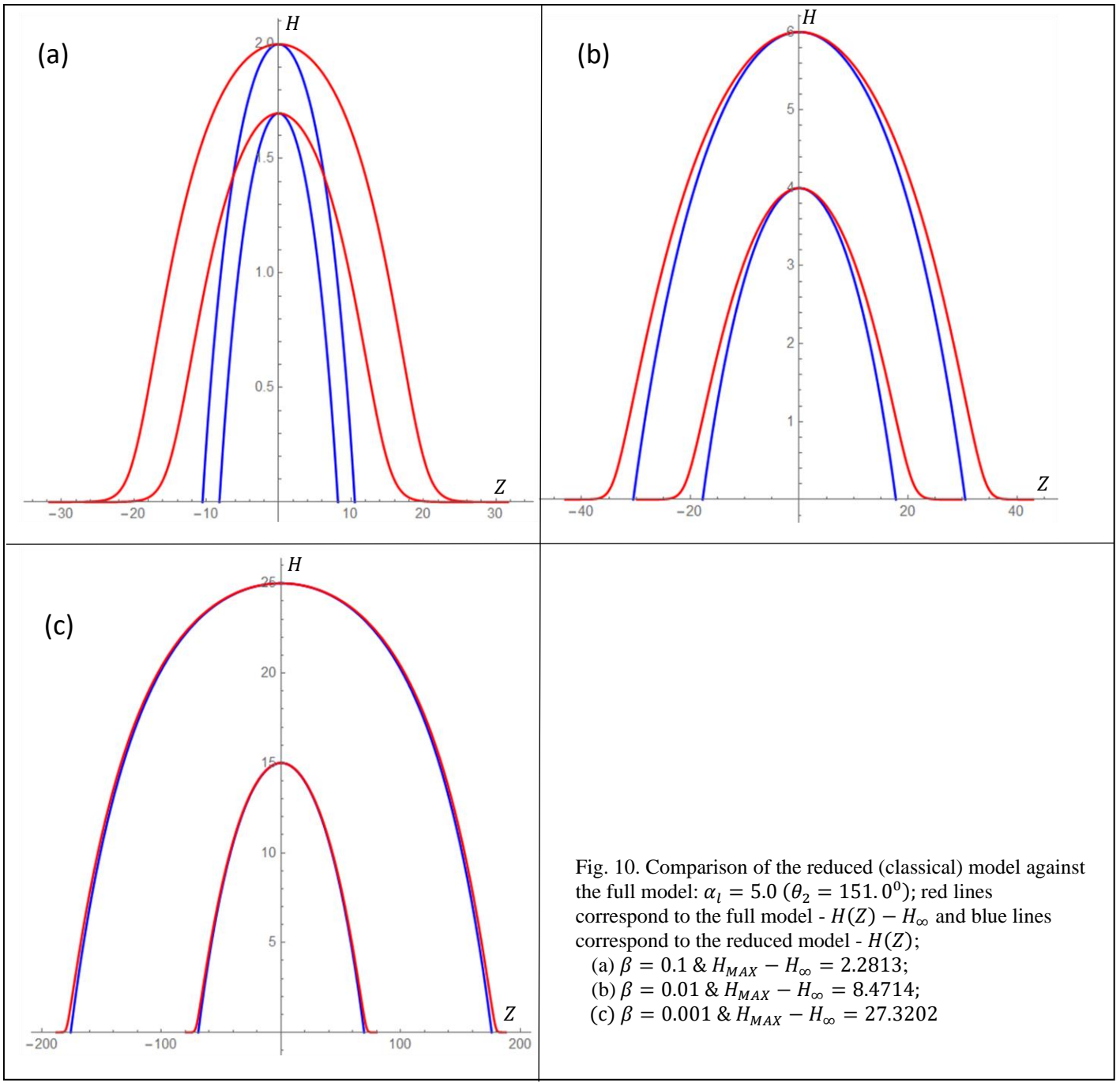

Fig. 10. Comparison of the reduced (classical) model against the full model: $\alpha_l = 5.0$ ($\theta_2 = 151.0^0$); red lines correspond to the full model - $H(Z) - H_\infty$ and blue lines correspond to the reduced model - $H(Z)$;
(a) $\beta = 0.1$ & $H_{MAX} - H_\infty = 2.2813$;
(b) $\beta = 0.01$ & $H_{MAX} - H_\infty = 8.4714$;
(c) $\beta = 0.001$ & $H_{MAX} - H_\infty = 27.3202$

**Acknowledgements**

The author would like to express his sincere gratitude to David Pekker for his terrific suggestions, which helped the author in the process of conducting this research, to Sean Cahill for his very valuable comments, and also to Marlene McDonald, Matthew Aubrey, Addison LaRock, James Cole-Henry, and Daniel Barnett for helpful discussions. The author would like also to acknowledge the support of FujiFilm Dimatix for conducting this research.**Appendix: Derivation of Eq. (21)**

Let us present Eqs. (19) and (20) in the following form,

$$h_{MAX} = \sqrt{\frac{2\gamma\tau}{\rho g}}, \tag{A1}$$

$$\tan(\theta_e) = \frac{\sqrt{2\tau - \tau^2}}{1 - \tau}, \tag{A2}$$

$$\tau = \frac{(m-n)\chi h^*}{(m-1)(n-1)\gamma}. \tag{A3}$$

Solving Eq. (A2) for $\tau$ yields

$$(\tan(\theta_e))^2 = \frac{2\tau - \tau^2}{1 - 2\tau + \tau^2} \rightarrow$$

$$\rightarrow (1 - 2\tau + \tau^2)(1 - (\cos(\theta_e))^2) = (\cos(\theta_e))^2(2\tau - \tau^2) \rightarrow$$

$$\rightarrow 1 - 2\tau + \tau^2 - (\cos(\theta_e))^2 + 2\tau(\cos(\theta_e))^2 - \tau^2(\cos(\theta_e))^2 =$$

$$= 2\tau(\cos(\theta_e))^2 - \tau^2(\cos(\theta_e))^2 \rightarrow$$

$$\rightarrow 1 - 2\tau + \tau^2 - (\cos(\theta_e))^2 = 0 \rightarrow$$

$$\rightarrow (1 - \tau)^2 = (\cos(\theta_e))^2 \rightarrow$$

$$\rightarrow \tau = 1 - \cos(\theta_e). \tag{A4}$$

Substituting Eq. (A4) into Eq. (A1), we obtain Eq. (21).

**REFERENCES**

[1]   M. Schick, in Liquids at interfaces, Les Houches Session XLIII, edited by J. Charvolin, J.-F Joanny, and J. Zinn-Justin (Amsterdam: Elsevier), pp. 415-497, 1990.

[2]   J. Israelachvili, Intermolecular and Surface Forces (Academic Press: London) 1992.

[3]   B. V. Derjaguin, N. V. Churaeb, and V. M. Muller, "Surface Forces", Consultation Bureau, New York (1987).18